\documentclass[a4paper,11pt]{iopart}
\usepackage{graphicx}
\usepackage{cite}
\usepackage{color}
\usepackage{amsfonts}
\usepackage{srcltx}
\begin{document}

\title{Equal-time correlation function for directed percolation}
\author{I. Beljakov and H. Hinrichsen}
\address{Universit\"at W\"urzburg,
	 Fakult\"at f\"ur Physik und Astronomie\\
         D-97074 W\"urzburg, Germany}

\ead{ibeljakov@physik.uni-wuerzburg.de}

\begin{abstract}
We suggest an equal-time $n$-point correlation function for systems in the directed percolation universality class which is well defined in all phases and independent of initial conditions. It is defined as the probability that all points are connected with a common ancestor in the past by directed paths. 
\end{abstract}

\def\ex#1{\langle #1 \rangle}
\def\xrightarrow{\rightarrow}

\section{Introduction}

One of the most studied classes of continuous phase transitions far from equilibrium is directed percolation (DP)~\cite{Kinzel85}. This transition occurs e.g. in stochastic lattice models for epidemic spreading, where infected sites either recover or infect a randomly selected nearest neighbor. Depending on the rates for infection and recovery, an initial epidemics may spread or disappear. Once the infection becomes extinct, the system enters a so-called absorbing state from where it cannot escape. Both regimes, survival and extinction, are separated by a well-defined phase transition. Directed percolation is the simplest universality class of phase transitions into absorbing states~\cite{MarroDickman99,Hinrichsen00,Odor04,Lubeck04,Odor08a,HenkelEtAl08a} and plays a paradigmatic role similar to the Ising model in equilibrium statistical mechanics.

Critical phenomena can be studied by analyzing correlation functions. In DP an important example is the so-called \textit{pair-connectedness} function $c(\vec r_1,t_1; \vec r_2,t_2)$ which is defined as the probability to find a directed connected path from a lattice site located at position $\vec r_1$ and time $t_1$ to another site at position $\vec r_2$ and time $t_2>t_1$ (see e.g. \cite{HenkelEtAl08a}). The pair connectedness function is well defined below, at, and above the critical point and plays the role of a propagator in the field-theoretic formulation of DP. However, if one is interested in \textit{spatial} correlations at a given instance of time the pair connectedness function does not provide any information because it vanishes for $t_1=t_2$. This raises the question how a meaningful equal-time correlation functions can be defined in models with absorbing states. 

\begin{figure}[t]
\includegraphics[width=160mm]{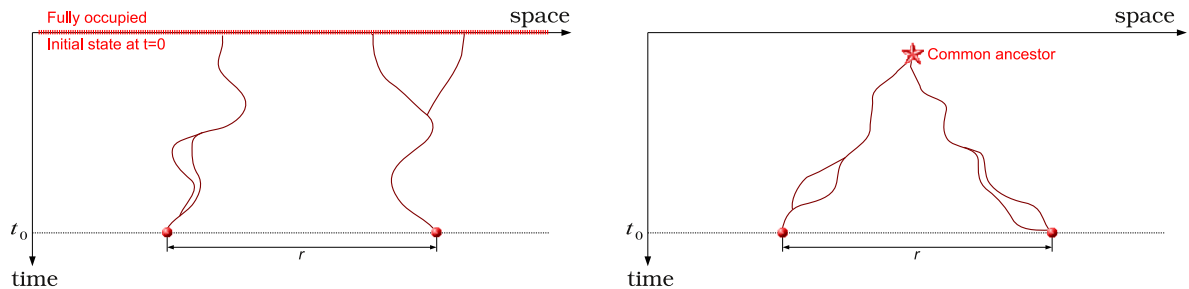}
\vspace{-5mm}
\caption{Equal-time two-point correlation functions. Left: Conventional density-density correlation function in 1+1 dimensions, defined as the probability that there are two directed paths connecting the initial state with the two points. Right: Correlation function investigated in the present work, defined as the probability that there exists a common ancestor in the past which is connected with both points by directed paths.}
\label{fig:sketch}
\end{figure}

The simplest equal-time correlation function, which has been studied in various contexts, is the density-density correlation function $\ex{s(\vec r_1,t) s(\vec r_2,t)}$, defined as the probability to find two sites at positions $\vec r_1$ and $\vec r_2$ simultaneously in the active (infected) state at time $t$ (see Fig.~\ref{fig:sketch}). This two-point function seems to be a natural candidate since it reminds us of a spin-spin correlation function in equilibrium statistical mechanics. Unfortunately this correlator depends on the history of the process. For example, if the system has not yet reached a stationary state, its value will depend on $t$ and the specific choice of the initial state. As a way out, one could measure the correlations in the stationary state, but then one is restricted to the active phase and the result will depend on the distance from criticality $\Delta=p-p_c$. To study the critical properties of the correlator, one has to keep the system in a slightly supercritical stationary state $0<\Delta \ll 1$ and to restrict the analysis to distances below the correlation length $r \ll r_c\sim \Delta^{-\nu_\perp}$. This method is of course numerically inefficient and is further complicated by crossover effects.

To overcome this problem, Dickman et. al.~\cite{DickmanOliveira05} devised a method that allows one to study equal-time correlations at the critical point. Starting point is the observation that the density of active sites in a finite system of lateral size $L$ at criticality first decays as usual, then stays for a while in a quasi-stationary state with a density $\rho\sim L^{\beta/\nu_\perp}$, until it suddenly enters the absorbing state. The key idea is to prevent the system from entering the absorbing state. This is done by keeping a list of typical configurations sampled in the quasi-stationary state and to reimpose one of them whenever the system would have entered the absorbing state, thereby artificially keeping it in its lowest excited state. Although this method became one of the standard tools in the study of absorbing phase transition, it is not entirely clear how the artificially stabilized state has to be interpreted. Moreover, the method involves finite-size effects which have to be handled with care. 

In this paper we propose a different type of equal-time correlation function which circumvents these difficulties. This correlator is defined as the probability that two points located at $\vec r_1$ and  $\vec r_2$ at equal time have a \textit{common ancestor somewhere in the past}, meaning that this ancestor is connected to both points by directed paths (see right panel of Fig.~\ref{fig:sketch}). Note that it only matters whether this ancestor exists, but not where it is located in space and time. Obviously, this definition can be generalized easily to $n$-point functions. 

Compared to the conventional density-density function, the correlator proposed here has several advantages. As it is defined in terms of paths, it is by definition history-independent and well-defined in the limit $L \to \infty$. Moreover, it can be analyzed everywhere in the phase diagram so that the trick by Dickman et. al. is no longer needed.

The main motivation for the present work came actually from a different direction. In equilibrium statistical mechanics the critical systems are usually not only scale-invariant but also conformally invariant. Especially in two dimensions this symmetry restricts the possible universality classes strongly, leading to a finite classification scheme of such transitions. There have been various attempts to generalize this concept to non-equilibrium systems by introducing an anisotropic version of conformal invariance (see e.g.~\cite{HenkelEtAl10a} and references therein). This concept was successfully applied to a large number of models but seemed to fail in the case of DP~\cite{Hinrichsen08a,Hinrichsen08b}. The ongoing debate triggered the question whether time-independent correlations of a two-dimensional DP process would exhibit not only scale-invariant but also conformal invariant features. This motivated us to search for a well-defined equal-time $n$-point function in DP, where the predictions of conformal invariance can be tested. At first glance it turned out that the correlation function defined above behaves like a conformally invariant one. However, extensive simulations reveal small violations.

\section{Numerical implementation}

\begin{figure}[t]
\includegraphics[width=160mm]{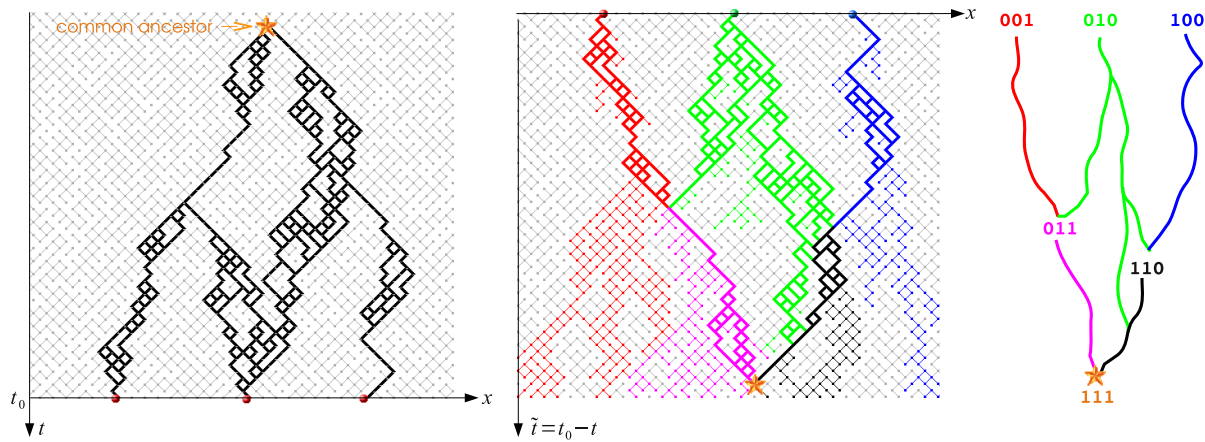}
\vspace{-5mm}
\caption{Example of a three-point function in 1+1 dimensions. Left: Configuration of open and closed bonds, where the three points marked by red dots have a common ancestor. Middle: The same configuration turned upside down may be interpreted as a time-reversed DP process starting with three active sites. The point where the three generated clusters merge for the first time corresponds to the common ancestor in the original setting. Right: The merging point is determined most easily by percolating bit patters (see text).}
\label{fig:configs}
\end{figure}

The equal-time $n$-point function $\Phi(\vec r_1,\ldots,\vec r_n)$ introduced in this paper is defined as the probability that there exists a lattice site in the past, a so-called common ancestor, which is connected to all points by directed paths.

Let us first discuss the question how this correlator can be estimated efficiently in a numerical simulation. To this end one has to choose a realization of DP on a lattice. For reasons to be explained below, we chose directed bond percolation, where the bonds of a diagonal lattice are open with probability $p$ and closed otherwise (see e.g.~\cite{MarroDickman99,Hinrichsen00,Odor04,Lubeck04,Odor08a,HenkelEtAl08a}). Activity percolates through open bonds along a preferred direction denoted as time. 

To estimate the correlator defined above, we could generate a spatio-temporal configuration of open and closed bonds, select $n$ points with given distances and trace along the open paths backwards in time to see whether all points have a common ancestor. This is shown in the left panel of Fig.~\ref{fig:configs} in the example of a three-point function in 1+1 dimensions. However, since the ancestor can be located anywhere, no matter how far it goes back time, any spatial or temporal restriction of the generated configuration of open and closed bonds will inevitably lead to finite-size effects.

A more efficient method exploits the so-called rapidity reversal symmetry of DP which states that directed paths forward and backward in time have the same statistical weight. This allows us to reverse the direction of time, use the $n$ points as initial seeds of activity, and search for a common descendant in the future. Note that the rapidity reversal symmetry is exactly fulfilled in the case of directed bond percolation, which explains why we chose this particular realization of DP.

To find out whether $n$ points have a common descendant, imagine that each seed carries a color, as shown in the central panel of Fig.~\ref{fig:configs}. By definition the common descendant is a lattice site where all these colors mix for the first time. In a numerical simulation this intuitive color scheme can be implemented by associating a string of $n$ bits with each lattice site and assigning one of these bits to each of the seeds (see right panel of Fig.~\ref{fig:configs}). These bit patterns then percolate forward in time along the open bonds. If two different patterns arrive at the same site, they coagulate by an OR operation. The first site carrying the bit pattern $11\ldots 1$ is a common descendant, corresponding to the common ancestor in the original situation. In this case the subroutine would return a `1', indicating that a common ancestor exists. However, if one of the bits dies out before a common descendant has been identified, the simulation stops and returns a `0'. The actual value of the correlator is then estimated by averaging over many independent runs.

Technically, instead of using static arrays, it is advantageous to keep the coordinates of sites with a non-vanishing bit pattern in a dynamically generated list without restricting the range of the coordinates. This technique accelarates the simultion significantly and eliminates possible finite-size effects. Moreover, it is useful to perform the simulation without a cutoff time until a `0' or a `1' is returned, which eliminates possible finite-time effects as well. 

\section{Scaling properties}
\label{sec:scaling}

The phenomenological scaling theory of absorbing phase transitions involves four independent critical exponents $\beta,\beta',\nu_\perp,\nu_\parallel$ for the order parameter and its reponse field as well as the spatial and temporal correlation lengths~\cite{HenkelEtAl08a}. In DP, the rapidity reversal symmetry implies that $\beta=\beta'$, leaving three of the bulk exponents independent. Moreover, it is useful to define the derived exponents
\begin{equation}
\chi=\beta/\nu_\perp\,,\qquad
\delta=\beta/\nu_\parallel\,\qquad
z=\nu_\parallel/\nu_\perp\,.
\end{equation}
In $d<4$ spatial dimensions the values of the critical exponents are not known exactly. Numerical estimates are listed in Table~\ref{tab:exp}.

\begin{table}
\begin{flushright}
\begin{small}\begin{tabular}{|c||c|c|c|c|c|c|} \hline
$d$ & $\beta$ & $\nu_\perp$ & $\nu_\parallel$ & $z=\nu_\parallel/\nu_\perp$ & $\chi=\beta/\nu_\perp$ & $\delta=\beta/\nu_\parallel$\\ \hline
1 & 0.276486(8) & 1.096854(4) & 1.733847(6) & 1.580745(10) & 0.252072(12) & 0.159464(6)\\
2 & 0.5834(30) & 0.7333(75)   & 1.2950(60)  & 1.7660(16)   & 0.7955(30)   & 0.4505(10)\\
\hline
\end{tabular}          \end{small}
\end{flushright}\label{tab:exp}
\caption{Numerical estimates of the DP critical exponents in $d$+1 dimensions, see~\cite{HenkelEtAl08a} and references therein.}
\end{table}

Generally, an $n$-point correlation function involves $n$ fields, each of them with a specific scaling dimension. In DP there are two primary fields, namely for creating and measuring activity. Because of the rapidity reversal symmetry they carry the same scaling dimension $[x]^{-\chi}$, where $[x]$ denotes the unit of length. For example, the pair connectedness function $c(r,t)$ mentioned in the introduction involves two fields and therefore carries the dimension $[x]^{-2\chi}$ \ (or equivalently $[t]^{-2\delta}$, where $[t]=[x]^z$ denotes the unit of time). This implies that for $r=0$ the pair connectedness function has to decay as $c(0,t)\sim t^{-2\delta}$. 

\begin{figure}[t]
\centering\includegraphics[width=150mm]{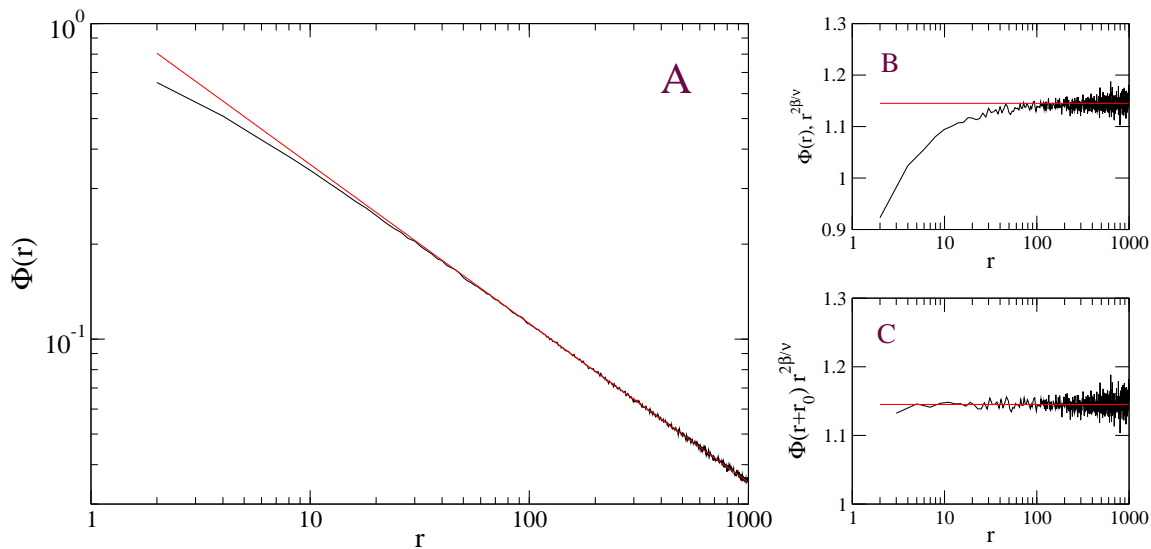}
\caption{\textbf{A}: Numerically determined two-point function $\Phi(r)$ in 1+1 dimensions (black) compared with the expected power law $r^{-2\chi}$ (red). \textbf{B}: Same data multiplied by $r^{-2\chi}$. \textbf{C}: Low distance correction (see Sect.~\ref{sec:scaling}).}
\label{fig:twopoint}
\end{figure}

The correlation function $\Phi(\vec r_1,\ldots,\vec r_n)$ considered in the present work depends on $n$ points and involves the common ancestor as an additional point. However, it is reasonable to assume that this point does \textit{not} contribute to the scaling dimension of the correlator because its location in space and time is arbitrary, meaning that it is effectively integrated out. Therefore, the scaling dimension of this correlator is $[x]^{-n\chi}$. For example, the two-point function is expected to decay as
\begin{equation}
\label{eq:twopoint}
\Phi(\vec r_1,\vec r_2) \;\simeq \; C\,|\vec r_1-\vec r_2|^{-2\chi}\,,
\end{equation}
where $C$ is a non-universal amplitude. As shown in Fig.~\ref{fig:twopoint}, this power-law decay is accurately reproduced in numerical simulations with $C=1.145(25)$.

The three-point function $\Phi(\vec r_1,\vec r_2, \vec r_3)$ carries the dimension $[x]^{-3\chi}$ and depends on three distances $r_{12},r_{13},r_{23}$, where $r_{ij}=|\vec r_i -\vec r_j|$. Therefore, it is expected to scale as
\begin{equation}
\label{eq:threepoint}
\Phi(\vec r_1,\vec r_2, \vec r_3) \;\simeq\; 
(r_{12} r_{13} r_{23})^{-\chi} \, F\Bigl(\frac{r_{12}}{r_{13}},\frac{r_{23}}{r_{13}}\Bigr)
\end{equation}
with a universal scaling function $F$ that depends only on scale-invariant ratios of these distances. 

As usual, scaling laws are only valid in the asymptotic limit of large distances, while for short distances one expects non-universal corrections to occur. For example, the two-point function shown in Fig.~\ref{fig:twopoint} clearly deviates from the predicted power law over the first two decades of time. In fact, the correlator is predicted to diverge if $r=|\vec r_1-\vec r_2|$ goes to zero. On the lattice, however, $\Phi(0)=1$ is always finite.

We propose that these short-distance corrections can be taken into account in a lowest-order approximation by shifting the distance parameter by a constant, i.e. Eq.~(\ref{eq:twopoint}) is replaced by
\begin{equation}
\label{eq:correction}
\Phi(r) \;\simeq \; C (r+r_0)^{-2\chi}\,.
\end{equation}
where $r_0$ is a constant. This constant should be chosen in such a way that the short-time deviations are minimized, as demonstrated in Fig.~\ref{fig:twopoint}C, leading us to the estimates $r_0 \approx 1.05(10)$ in $d=1$ and $r_0 \approx 0.31(1)$ in $d=2$ spatial dimensions. Note that these values are expected to be non-universal, i.e. they may differ for different realizations of DP. Moreover, we would like to point out that the constant should not be derived from the condition $\Phi(0)=1$ since in this point higher-order corrections would lead to an incorrect estimate.

\section{Test of conformal invariance}

\begin{figure}[t]
\centering\includegraphics[width=110mm]{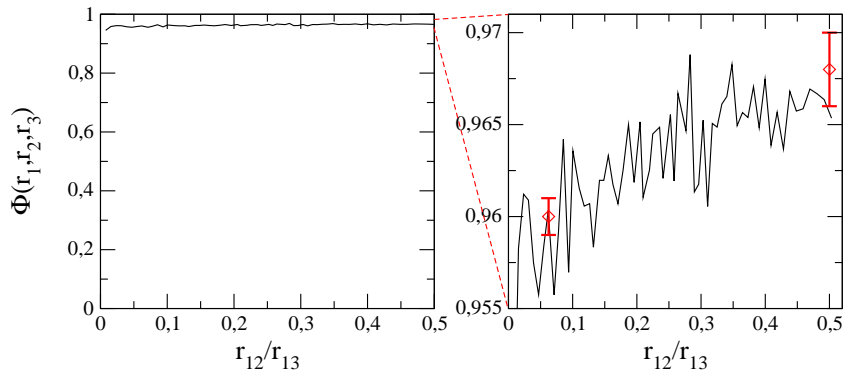}
\caption{Three-point correlation function in $d=1$ spatial dimension. The graphs show the value of the scaling function $F(\frac{r_{12}}{r_{13}},\frac{r_{23}}{r_{13}})=\Phi(r_{1},r_{2},r_{3}) /(r_{12}r_{13}r_{23})^{\chi}$ for $r_{13}=256$ fixed and $r_{12}$ varying between 0 and 128. The red dots with error bars report two precision measurements which confirm the drift. }
\label{fig:1d}
\end{figure}

In 1970, Polyakov argued that scale-invariant equilibrium systems should also be invariant under \textit{local} scale transformations~\cite{Polyakov70a}. This conformal invariance greatly restricts the form of correlation functions. In particular Polyakov showed that the three-point function at criticality should take the form
\begin{equation}
G_{III}(\vec r_1,\vec r_2,\vec r_3) \;=\; const \; r_{12}^{\Delta_3-\Delta_1-\Delta_2}\; r_{13}^{\Delta_2-\Delta_1-\Delta_3}\; r_{23}^{\Delta_1-\Delta_2-\Delta_3}\,,
\end{equation}
 where $r_{ij}=|\vec r_i-\vec r_j|$ and $\Delta_i$ is the scaling dimension of the field measured at position $\vec r_i$. This equation is expected to hold in any dimension. Later it was shown that conformal invariance in two dimensions, where the conformal group has infinitely many generators, is even more restrictive and leads to a complete classification scheme of equilibrium critical phenomena~\cite{Cardy96,Henkel99}.

In non-equilibrium statistical mechanics such a classification scheme is not yet known. Although there have been attempts to generalize conformal invariance to anisotropic and time-dependent systems~\cite{HenkelEtAl10a}, DP turned out to be incompatible with the predictions of such theories~\cite{Hinrichsen08a,Hinrichsen08b}. This led to the question of whether at least the time-independent properties of DP carry a signature of conformal invariance, motivating  us to define a history-independent equal-time correlator. According to Polyakov, the three-point correlator defined above with scaling dimensions $\Delta_1=\Delta_2=\Delta_3=\chi$) is conformally invariant if  
\begin{equation}
 \Phi(\vec r_1,\vec r_2,\vec r_3) \;=\; \frac{const}{(r_{12}r_{13}r_{23})^{\chi}}\,, 
\end{equation}
meaning that the scaling function $F$ in Eq.~(\ref{eq:threepoint}) is constant.

To test this prediction, we first consider the one-dimensional case, where the three points $\vec r_1,\vec r_2,\vec r_3$ lie on a straight line. In this case $\frac{r_{12}}{r_{13}}+\frac{r_{23}}{r_{13}}=1$ so that the scaling function depends only on one argument, say $\frac{r_{12}}{r_{13}}$. The numerical results are shown in Fig.~\ref{fig:1d}. Plotted on a scale between 0 and 1 (left panel), the scalling function seems to be almost constant. However, the magnification of the data shown in the right panel exhibits a clear systematic drift of roughly one percent. High-precision simulations at individual points confirm that this drift does not go away as one increases the distances even further, meaning that it cannot be attributed to lattice- or finite-size effects.

To investigate the correlation function in 2+1 dimensions we performed simulations using a fixed lattice with $1024\times 1024$ sites and periodic boundary conditions. In order to increase the numerical efficiency, we implemented a parallel version of the algorithm based on the programming language CUDA developed by NVidia, which allows one to access the multi-processor GPU of a graphics adapter. The point $\vec r_1$ was fixed during the simulation, while the locations of the second and the third point varied in the two-dimensional plane within the radius of 5 (border to the higher-order lattice effects ) to 90 lattice sites from the first point. At these relatively small distances we expect finite size effects to be negligible.

\begin{figure}[t]
\includegraphics[width=80mm]{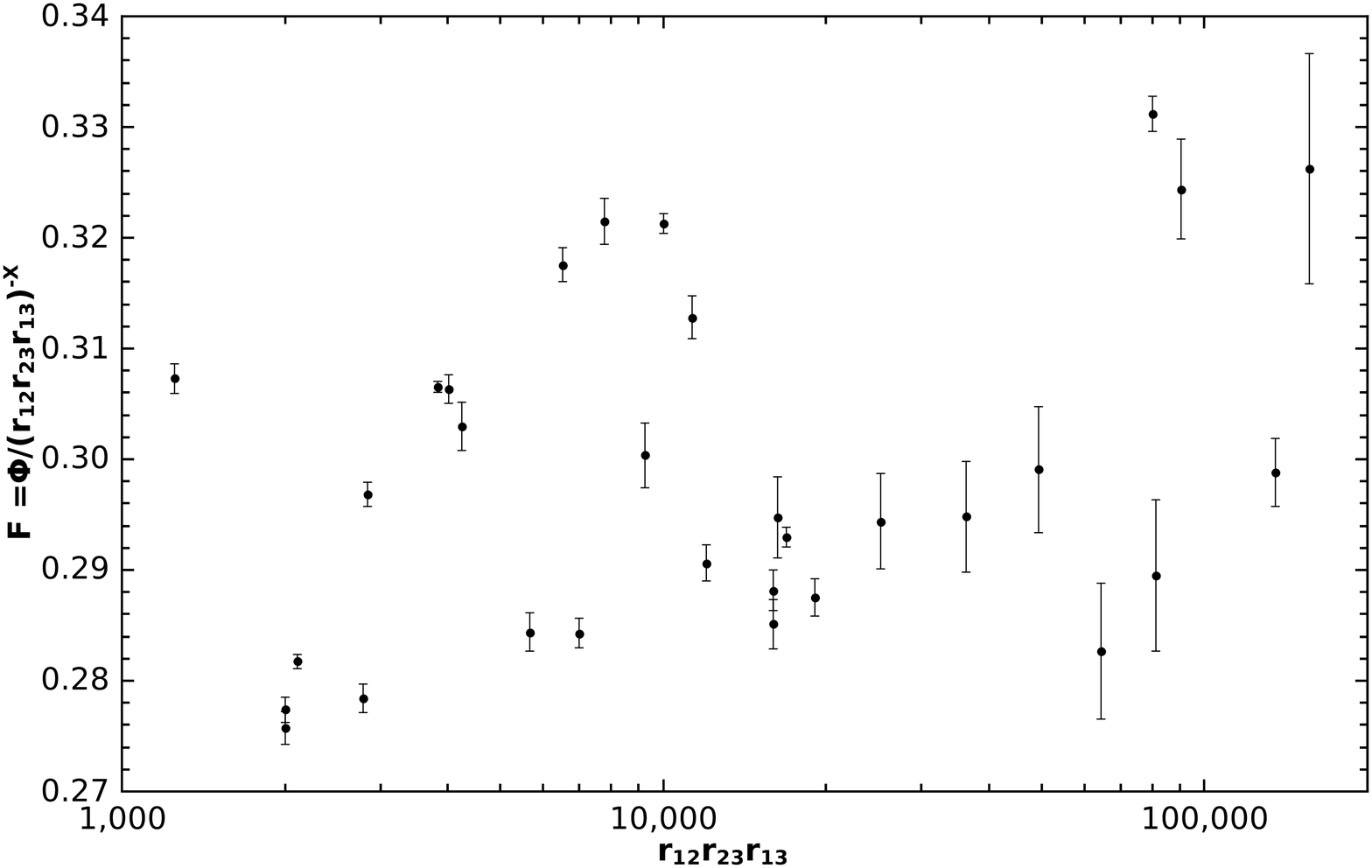}
\includegraphics[width=80mm]{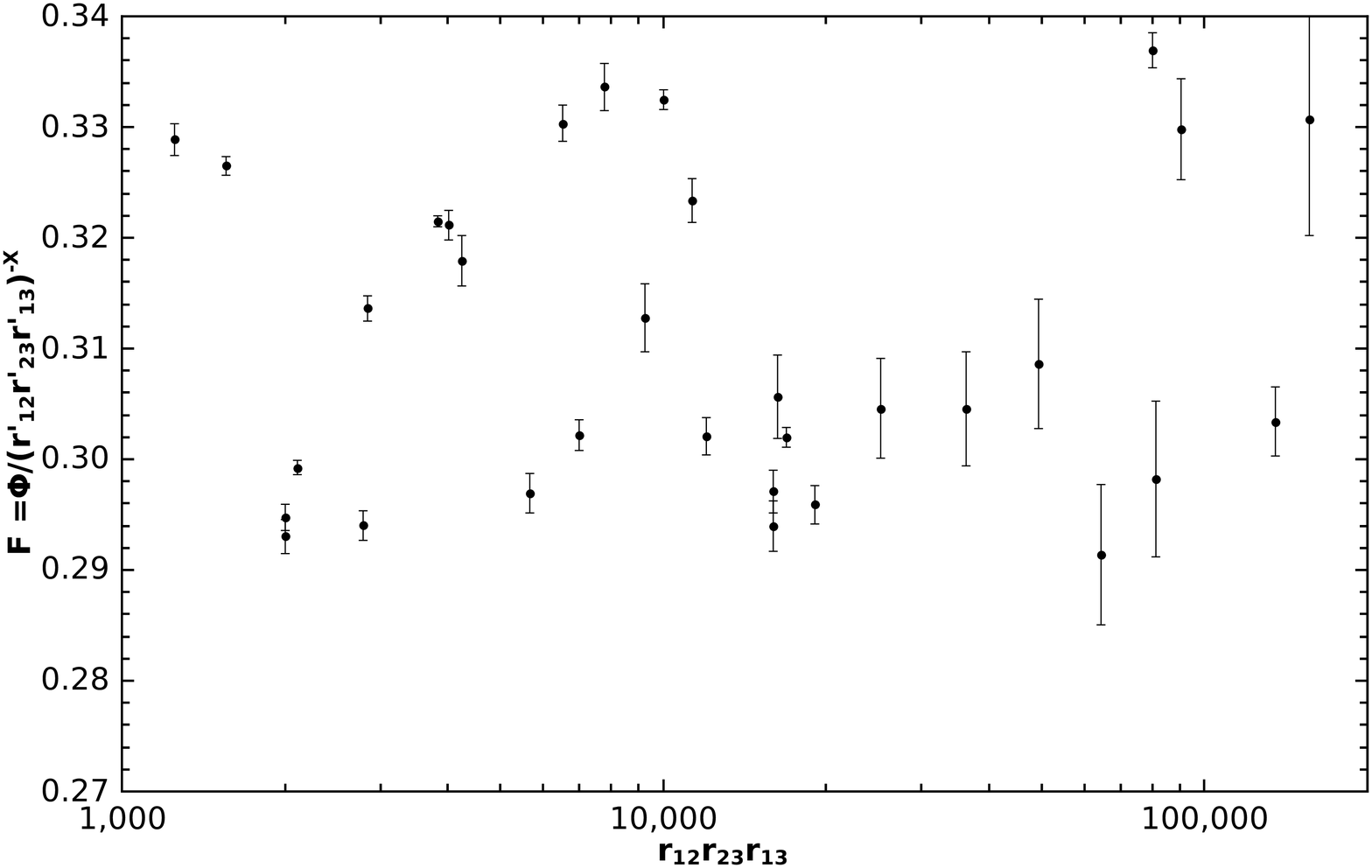}
\vspace{-5mm}
\caption{Scaling function $F(\frac{r_{12}}{r_{13}},\frac{r_{23}}{r_{13}})$ of the three-point function in 2+1-dimensions. The distances $r_{12}$, $r_{13}$, and $r_{23}$ vary between 5 and 90 lattice sites. The results are shown on the left side in a one-dimensional projection by plotting $F$ as a function of $r_{12}r_{13}r_{23}$. The right panel shows the same data including short-distance corrections according to Eq.~(\ref{eq:correction}). The error bars refer to the statistical error only, they do not include possible systematic errors.}
\label{fig:simul2d}
\end{figure}

The results shown in Fig.~\ref{fig:simul2d} confirm the findings in one dimension: The scaling function is finite and has no signatures of zeros or singularities. It varies more strongly than in the previous case but still in a range of less than $\pm 10 \%$.

\section{Strongly Asymmetric limit}

In statistical physics most scaling function show a singular behavior, i.e. they vanish or diverge if one their parameters goes to zero or infinity. The scaling function $F$, however, remains finite even when two of the three points approach each other. As we will show in the following, the correlation function can be factorized in this limit.

\begin{figure}[t]
\centering\includegraphics[width=100mm]{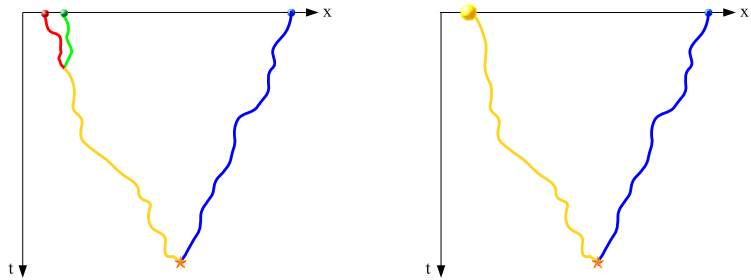}
\vspace{-3mm}
\caption{Left: Anisotropic limit of a three-point function in the time-reversed setting. In practically all configurations, where the three points have a common descendant, the colors of the nearby points mix already after very short time, thus effectively behaving like a single seed (right panel).}
\label{fig:factorize}
\end{figure}

To understand this factorization intuitively let us consider two clusters generated two nearby points in the time-reversed picture (e.g. the red and the green one in Fig.~\ref{fig:factorize}). Generating such clusters and selecting those which survive for a long time one observes that almost all of them consist mainly of a single color (red, green or its mixture), while the coexistence of different colors is exponentially suppressed for large $t$. Only clusters surviving with a mixture (symbolized by the orange line in Fig.~\ref{fig:factorize}) can contribute to the three-point function if they merge with the cluster generated by the distant point.

From the perspective of the common descendant the points with small distance on the left side behave effectively like single seed (see right panel) with a certain survival probability. Knowing this survival probability and the two-point function, this allows one to predict the three-point function in the limit that two of the three points are very close.

More specifically, let $r=r_{12}$ be the small distance and $R\approx r_{13}\approx r_{23}$ the mean of the large distances between the particles so that we can express the three-point function $\Phi(R,r)$ as a function of these two distances. Moreover, let $P_1(t)$ be the usual survival probability of a cluster starting with a \textit{single} seed and $P_2(r,t)$ be the survival probability that a cluster starting with \textit{two} seeds of different color at distance $r$ will survive in a mixed-colored state. Dimensional power counting implies the scaling laws
\begin{equation}
  \label{P_2}
  P_1(t)\simeq A \, t^{-\delta}\,, \qquad P_2(r,t) \simeq B \, r^{-\chi} t^{-\delta}
\end{equation}
with certain non-universal amplitudes $A$ and $B$. The factorization conjecture is based on the assumption that for $r\ll R$ the three-point function $\Phi(R,r)$ is proportional to the two-point function $\Phi(R)$ because the pair of nearby particles can be replaced by an effective single seed. However, the survival probability of a pair in a mixed state differs from the ordinary survival probability of a single seed, meaning that the proportionality factor is just $P_2(r,t)/P_1(t)$, hence
\begin{equation}
\Phi(R,r) \;\simeq\; \frac{P_2(r,t)}{P_1(t)}\Phi(R) \;\simeq\; \frac{BC}{A} r^{\chi} R^{-2\chi}
\qquad\mbox{for } r\ll R\,.
\end{equation}
Since the time dependence cancels out, this result is compatible with the scaling form~(\ref{eq:threepoint}) with the special value of the scaling function
\begin{equation}
F(0,1)=\frac{BC}{A}\,.
\end{equation}
Determining the constants $A,B,C$ numerically the values of this expression are compatible with the values obtained in direct numerical simulations (see Table~\ref{tab:F}). This confirms that the scaling function does not diverge, even if two of the three points approach each other. Moreover, the numerical values support  the observation that the scaling function is not constant but varies in a narrow range.

\begin{table}
\begin{small}\begin{center}
\begin{tabular}{|c||c|c|c|c|c|c|} \hline &&&&&&\\[-4mm]
$d$ & $F(1,1)$ & $F(\frac12,\frac12)$ & $A$ & $B$ & $C$ & $F(0,1)=\frac{BC}{A}$  \\
 \hline
1 & impossible & 0.968(5) & 0.9325(5) & 0.78(1) & 1.145(25) & 0.957(10)\\
2 & 0.334(2) & 0.294(2) & 0.9464(16) & 0.36(1) & 0.752(2) & 0.286(9)
\\ \hline
\end{tabular}
\end{center}
\end{small}
\caption{Special values of the scaling function (\ref{eq:threepoint}) for an equilateral triangle $r_{12}=r_{13}=r_{23}$, three points on a line with equal distances $r_{12}=r_{23}=\frac{r_{13}}{2}$, and in the limit $r_{12}\ll r_{13}\approx r_{23}$.}
\label{tab:F}
\end{table}

\begin{figure}[h]
\centering\includegraphics[width=100mm]{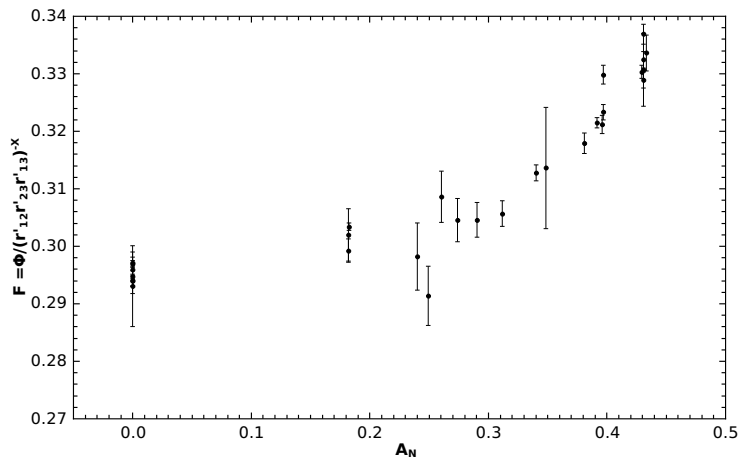}
\vspace{-3mm}
\caption{Scaling function $F$ with short-range corrections as a function of the normalized area $A_N$ (see Eq.~(\ref{eq:a_n})). $A_N = 0$ means that the three points lie on a straight line, and $A_N = 0.433$~(maximum value) represents an equilateral triangle.}
\label{fig:area}
\end{figure}

In Table~\ref{tab:F} we can see, that an equilateral triangle provides the higher value of $F$, than a straight line. Considering other constellations we saw that all other values of $F$ vary between these two. We plotted $F$ as a function of so-called normalized area of the triangle   
\begin{equation}
A_N = A_{triangle}/(r_{12} r_{23} r_{13})^{2/3}
\label{eq:a_n}
\end{equation}
and could recognize a certain correlation between these variables (see Fig.~\ref{fig:area}). Both of them do not exceed a certain range: $F \in [0.294(2);0.334(2)]$ and $A_N\in[0,\sqrt{3}/4]$, and $F$ grows with $A_N$. Although it was not possible to determine an exact function due to a large error range, this observation can be helpful for further study of the scaling function.

\section{Conclusions}

In this work we have introduced a novel kind of equal-time correlation function for directed percolation which is defined as the probability that $n$ points have a common ancestor in the past, no matter where this ancestor is located. This correlation function is well-defined in the whole phase diagram and does not depend on the history of the process. Using the rapidity reversal symmetry, this correlation function can be related to the probability that the clusters generated by $n$ seeds will merge at some point in the future, which allows one to compute the correlator efficiently.

Analyzing the scaling behavior of the three-point function we find that the universal scaling function is almost constant, which would be an indicator of conformal invariance. However, extensive simulations reveal a small variation of (in one dimension) less than one percent. 

The bottom line of this paper is a warning: Some properties of directed percolation behave almost as if the process was conformally invariant, but high-precision simulations reveal that this is not the case. This means that numerical results can easily generate misleading expectations and have to be interpreted with care.

\section*{References}


\begin{thebibliography}{10}
\expandafter\ifx\csname url\endcsname\relax
  \def\url#1{{\tt #1}}\fi
\expandafter\ifx\csname urlprefix\endcsname\relax\def\urlprefix{URL }\fi
\providecommand{\eprint}[2][]{\url{#2}}

\bibitem{Kinzel85}
Kinzel W 1985 {\em Z. Phys. B\/} {\bf 58} 229

\bibitem{MarroDickman99}
Marro J and Dickman R 1999 {\em {Nonequilibrium phase transitions in lattice
  models}\/} (Cambridge, UK: Cambridge University Press)

\bibitem{Hinrichsen00}
Hinrichsen H 2000 {\em Adv. Phys.\/} {\bf 49} 815 [cond-mat/0001070]

\bibitem{Odor04}
\'Odor G 2004 {\em Rev. Mod. Phys.\/} {\bf 76} 663

\bibitem{Lubeck04}
L{\"u}beck S 2004 {\em Int. J. Mod. Phys. B\/} {\bf 18} 3977

\bibitem{Odor08a}
{\'O}dor G 2008 {\em Universality in nonequilibrium lattice systems\/}
  (Singapore: World Scientific)

\bibitem{HenkelEtAl08a}
Henkel M, Hinrichsen H and L{\"u}beck S 2008 {\em Non-Equilibrium phase
  transitions\/} vol~1 (Berlin, Germany: Springer)

\bibitem{DickmanOliveira05}
Dickman R and {de~Oliveira} M~M 2005 {\em Phyisca A\/} {\bf 357} 134

\bibitem{HenkelEtAl10a}
Henkel M and Pleimling M 2010 {\em Non-Equilibrium phase transitions\/} vol~2
  (Berlin, Germany: Springer)

\bibitem{Hinrichsen08a}
Hinrichsen H 2008 {\em J. Stat. Mech.: Theor. Exp.\/}  P02016

\bibitem{Hinrichsen08b}
Hinrichsen H 2008 {\em J. Stat. Mech.: Theor. Exp.\/}  P07026

\bibitem{Polyakov70a}
Polyakov A~M 1970 {\em JETP Lett.\/} {\bf 12} 381

\bibitem{Cardy96}
Cardy J 1996 {\em {Scaling and renormalization in statistical physics}\/}
  (Cambridge, U.K.: Cambridge University Press)

\bibitem{Henkel99}
Henkel M 1999 {\em {Conformal Invariance and Critical Phenomena}\/} (Berlin:
  Springer)

\end{thebibliography}

\providecommand{\newblock}{}

\end{document}